\documentstyle[12pt]{article}
\def\beq{\begin{equation}}
\def\eeq{\end{equation}}
\def\bea{\begin{eqnarray}}
\def\eea{\end{eqnarray}}
\def\bq{\begin{quote}}
\def\eq{\end{quote}}

\makeatletter
\def\vereq#1#2{\lower3pt\vbox{\baselineskip1.5pt \lineskip1.5pt
\ialign{$\m@th#1\hfill##\hfil$\crcr#2\crcr\sim\crcr}}}
\makeatother

\begin{document}

\begin{titlepage}
\begin{center}
\today     \hfill    SLAC-PUB-8008\\
~{} \hfill SU-ITP-98/62\\
~{} \hfill hep-ph/9811353\\

\vskip .05in

{\large \bf
New origin for approximate symmetries from \\ 
distant breaking in extra dimensions}
\vskip 0.05in

Nima Arkani-Hamed$^a$ and Savas Dimopoulos$^b$

\vskip .02in
{\em $^a$ SLAC, Stanford, California 94309, USA}
\vskip 0.05truecm
{\em $^b$ Physics Dept., Stanford University,
Stanford, California 94309, USA}
\end{center}

\begin{abstract}
The recently proposed theories with TeV-scale quantum
gravity do not have the usual ultraviolet
desert between $\sim 10^{3} - 10^{19}$ GeV where effective
field theory ideas apply. Consequently, the success of the
desert in explaining approximate symmetries is lost, and
theories of flavor, neutrino masses, proton longevity or
supersymmetry breaking, lose their usual habitat.
In this paper we show that these ideas can find a new home in
an infrared desert: the large space in the extra dimensions.
The main idea is that symmetries are primordially exact on our brane,
but are broken at $O(1)$ on distant branes. This breaking is communicated
to us in a distance-suppressed way by        
bulk messengers. We illustrate these ideas in a number of
settings: 1) We construct theories for the fermion mass hierarchy which
avoid problems with large flavor-changing neutral currents.
2) We re-iterate that proton stability can arise if  
baryon number is gauged in the bulk.
3) We study limits on light gauge fields and scalars in the bulk coming
from rare decays, astrophysics and cosmology.
4) We remark that the same ideas can be used to explain small
neutrino masses, as well as hierarchical supersymmetry breaking.   
5) We construct a theory with bulk technicolor, 
avoiding the difficulties with extended technicolor.
There are also a number of interesting experimental signals of these
ideas: 1) Attractive or repulsive, isotope dependent
sub-millimeter forces $\sim 10^6$ times gravitational strength, from
the exchange of light bulk particles.
2) Novel Higgs decays to light generation fermions plus
bulk scalars.
3) Collider production of bulk vector and scalar fields, 
leading to $\gamma$ or jet+ missing energy signals as in the case
of bulk graviton production, with comparable or larger rates.

\end{abstract}

\end{titlepage}
\section{Life without the desert}
The standard paradigm of particle physics dates back to the advent of
grand unified theories
\cite{gut} or perhaps even further back to Fermi's theory of beta decay.
Its premise is that there are two fundamental scales
--the weak and Planck masses-- separated by a
 large ``desert".  The existence of the desert plays a fundamental
role in formulating and solving problems in particle physics.
Examples include the physics of flavor, neutrino masses and of
unification. Much of the
physics of the early universe takes place when the temperature of the
universe is in the desert.
The very hierarchy problem is simply the statement of the large size of
the desert. This suggested a
new proposal for solving the
hierarchy problem simply postulating that the
desert does not exist: namely that the
fundamental scale of gravity is
in fact identical to the weak interaction scale
$\sim$ TeV \cite{ADD1,AADD,ADD2}. In this new paradigm, the observed
weakness of gravity at long distances is due the
existence of new sub-millimeter spatial
dimensions into which gravity spreads. The
standard model fields are localized to a
$(3+1)$-dimensional wall or ``3-brane''. The
hierarchy problem becomes isomorphic to the
problem of the large size of the extra
dimensions. Some ideas for stabilizing large dimensions have been
explored in \cite{raman2, radus}.

Whereas the absence of a desert may allow for a
novel approach to the hierarchy problem, it also
deprives us from all mechanisms whose existence
relied on the desert. This includes baryon
stability, naturalness of approximate lepton
number conservation and neutrino masses, as well
as approximate neutral flavor conservation which
are some of the successes of the
{\it non-supersymmetric} standard model
\footnote{Note that the naturallness of these successes of
the standard model is lost in any $\sim$ TeV
extension of the standard model, including low
energy supersymmetry}. In addition there are
phenomena and mechanisms in extensions of the
standard model such as the supersymmetric gauge
coupling constant unification, electroweak
breaking in technicolor or susy, supersymmetry
breaking as well as models of flavor which
largely relies on the existence of the desert.
It is the purpose of this paper to find new and
natural mechanisms to account for some of these
phenomena within the new framework of theories
without a desert.

While we have been deprived of the desert in short-distance scales
between the weak and Planck scales, we have gained the
large space in the extra dimensions where new mechanisms can reside.
In this paper we present ideas for generating natural flavor hierarchies
of quark and
lepton masses and show that they do not lead to unacceptably large
flavor violations.
The same generic mechanism can be used to understand the approximate
conservation of baryon and
lepton number. We also note in passing that in theories with a high
fundamental
Planck scale, these same ideas can furnish a simple mechanism for
dynamical
supersymmetry breaking.

The basic idea which makes all of this possible
is that symmetries which are primordially good
on our brane maybe broken on other distant branes by a large
$O(1)$ amount. The information of this breaking is transmitted to us by
messenger fields living in the bulk,
and is suppressed by the distance between our brane and the others. If
the messengers are massive, an exponential
suppression of the symmetry breaking on our wall results, while even if
they are very light a power suppression is
possible.

The case of light messengers is particularly interesting since they can
macroscopic forces which
may be detected by the upcoming sub-millimeter tests of gravitational
strength forces \cite{Price}. Indeed,
in many cases, the forces are expected to be at least $10^{6}$ times
gravitational strength.
Of course, the constraints on very light messengers coming from rare
decays, astrophysics and cosmology must also
be considered. We find that there is still a significant region of
experimentally viable parameter space where the
dramatic predictions of $10^6$ times gravity sub-millimeter forces are
possible.

If the bulk messengers are heavier than a few GeV, all the previous
constraints disappear, as does the signal for
sub-millimeter experiments. On the other hand, in some cases, novel
Higgs decays to light SM generations
and bulk modes are possible, which can be comparable or even dominate
over the
usual decay channels. Furthermore, the production of bulk gauge
fields or scalars can be probed via $\gamma$ or jet + missing energy
signals
much as in the case of bulk graviton production, but with a much larger
rate.

Finally, as a somewhat different illustration of new possibilities
for interesting phenomenology from extra dimensions, we show that
electroweak symmetry breaking on our wall does not require
a fundamental Higgs field, but can be induces via a technicolor gauge
force in the
bulk triggering a technifermion condensate on our wall. This can
be combined with the other ideas for generating flavor, thereby avoiding
the usual problems with extended technicolor.

\section{Yukawas from distant flavor breaking}
A popular mechanism for explaining the smallness of the observed Yukawa
couplings
invokes a flavor symmetry $G_F$ under which the SM generations, and
perhaps the Higgs, are charged.
The representations under $G_F$ are chosen such that none of the light
generation Yukawa couplings
$f f^c H$ are neutral under $G_F$, but the large Yukawa couplings
are invariant. In addition, there
are a certain number of ``flavon" fields $\phi$, charged under
$G_F$, which are assumed to acquire vevs from some dynamics thereby
breaking $G_F$. In the effective theory beneath some scale
$M_F$, we expect $G_F$ invariant higher dimension operators of the
generic form
\begin{equation}
{\cal O} \sim \lambda \frac{\phi^{n}}{M_F^n} f f^c H
\label{FN}
\end{equation}
with $\lambda \sim O(1)$ to be present in the theory. If $M_F$ is close
to the string scale $M_s$  these operators can be generated by quantum
gravity effects.
If $M_F \ll M_s$ these operators can be generated by integrating out
heavy fields of mass $\sim M_F$ which carry flavor charge
(the Froggatt-Nielsen mechanism \cite{FN}). Regardless, if we assume
that
$\epsilon \sim \langle \phi \rangle /M_F$ is small, the
small Yukawa couplings can be understood as powers in the small parameter
$\epsilon$.
Of course, this does not constitute a full understanding of the
smallness of the fermion masses until the mechanism for
triggering a vev for $\phi$ smaller than $M_F$ is presented.

In this section, we will show how in the presence of extra dimensions
populated with multiple 3-branes, we can take advantage of
flavor symmetries to explain small Yukawa couplings without needing a
small flavor breaking scale $\langle \phi \rangle \ll M_F$.
In fact, we will assume that the flavor symmetry is badly broken at the
string scale $\langle \phi \rangle \sim M_s$. However, the flavons
$\phi$ are taken to live on a different 3-brane than ours.
Therefore, some ``messenger" needs to communicate the information of
flavor breaking to our
3-brane. It is easy to see that gravity in the bulk cannot generate any
$G_F$ violating operator
involving wall fields. The reason is simple:
gravity is not charged under $G_F$. Any operator generated by
gravitational exchange will be
of the form $O_{our wall} O_{other wall}$ where each $O$ is individually
$G_F$ symmetric.
This generalises to an intuitively
obvious statement: no $G_F$ violating operators can be induced on our
wall unless
there are bulk modes charged under $G_F$.
If this messenger is massive, the resulting coefficient $\lambda$ in
eqn.(\ref{FN}) will be exponentially suppressed by the distance between
the
3-branes, while even if the messenger is massless, a power suppression
of $\lambda$ is possible.

We note here that the generic mechanism for obtaining small Yukawas
presented below works for
any value of $M_s$, not just for $M_s$ near the TeV scale.
Nonetheless, we will mostly be interested in the latter case,
where the correlation between generating flavor while avoiding
flavor-changing problems is particularly challenging and
interesting.
It is easy to understand the idea in the following simple toy example.
Suppose that there is an abelian (continuous or discrete)
$G_F$ such that the electron Yukawa coupling
$L_1 H E^c_1$ carries flavor charge $1$, and that there is a flavon
$\phi$ with charge $-1$, which acquires a string scale vev
$\langle \phi \rangle \sim M_s$, but which lives on a different 3-brane
parallel to ours, a transverse distance $r$ away from our 3-brane.
In order to communicate the information of $G_F$ breaking between the
3-branes, we have a scalar field $\chi$ of mass $m_{\chi}$, with charge
$-1$ under $G_F$.
For definiteness, let the co-ordinates of
our 3-brane be $(x,y^a=0)$ while those of the other brane are
$(x',y^a=y_*^a)$, where $a=1,\cdots,n$ runs over the extra
``large" dimensions and $|y_*|=r$ is the transverse distance between the
branes.
We expect all possible local operators consistent with symmeties to be
generated by physics
above the string scale $M_s$, in particular
we can have
\begin{equation}
{\cal L} \supset
\int_{\mbox{us}} d^4 x LHE^c(x) \chi(x,y^a=0) +
\int_{\mbox{other}} d^4 x' \phi(x') \chi^*(x',y^a=y^a_*)
\end{equation}
where we have set $M_s = 1$.
For simplicity, let us for the moment ignore possible non-linear
self-interactions of the $\chi$ field with itself in the bulk.
Then, the non-zero vev if $\phi$ acts as a linear source,``shining"
the $\chi$ field. The profile of the $\chi$ field created by this
source is easily obtained. Since $\langle \phi \rangle$ is $x'$
independent, the value $\langle \chi \rangle$ only depends on the
transverse
distance from the other brane, and is given by the Yukawa
potential in the $n$ transverse dimensions:
\beq
\langle \chi \rangle(x,y) = \langle \phi \rangle \times
\Delta_n(|y-y_*|)
\eeq
where
\beq
\Delta_n(y) = \left(-(\partial^2)_n + m^2_\chi\right)^{-1}(y) =
\int d^n \kappa e^{i \kappa y} \frac{1}{\kappa^2 +
m_{\chi}^2}.
\label{delt}
\eeq
More explicitly, we have
\begin{eqnarray}
\langle \chi \rangle(x,y) &=& \int d^4 x' \left((\partial^2)_4 -
(\partial^2)_n + m_{\chi}^2 \right)^{-1}(x-x',y-y_*) \langle \phi
\rangle(x')
\nonumber \\
&=& \langle \phi \rangle \times \left(-(\partial^2)_n +
m^2_{\chi} \right)^{-1}(y-y_*).
\end{eqnarray}
Note that we should actually be using the propagator appropriate
to the compactified space (i.e. the integral over $\kappa$ in
eqn.(\ref{delt})
should be replaced with a sum over KK modes), however,
the difference is negligible as
long as $|y-y_*|$ is less than $r_n$, the size of the extra dimensions.

There is now a Yukawa coupling generated from
the ``shined" value of $\langle \chi \rangle$ on our wall
\begin{equation}
{\cal O} = \int d^4 x \langle \chi \rangle (x,y=0) LHE^c(x) = \int d^4 x
\langle \phi \rangle
\Delta_n(r) LHE^c(x).
\end{equation}
Recall that for $n>2$, we have
\begin{eqnarray}
\Delta_n(r) &\sim& \frac{1}{r^{n-2}} \, (r m_{\chi} \ll 1) \nonumber \\
&\sim& \frac{e^{-m_{\chi} r}}{r^{n-2}} \,(r m_{\chi} \gg 1)
\end{eqnarray}
while for $n=2$
\begin{eqnarray}
\Delta_{n}(r) &\sim& -\mbox{log}(r m_{\chi}) \, (r m_{\chi} \ll 1)
\nonumber \\
&\sim& \frac{e^{-m_{\chi} r}}{\sqrt{m_{\chi} r}} \, (r m_{\chi} \gg 1).
\end{eqnarray}
Thus, we get the expected $e^{-mr}$ suppression of the electron Yukawa
coupling
if the branes are separated by more than $m_{\chi}^{-1}$,
while for $n>2$, we still get the geometrical  power suppression $\sim
r^{2-n}$ even for $m_{\chi}=0$.
Of course,  we can not have a massless scalar coupled to the ordinary
stable matter
because of conflicts with the tests of the equivalence principle. In any
case, we do not expect scalars
to stay light unless protected by some symmetry. Interestingly, if there
is SUSY in the bulk with SUSY broken on the wall,
the scalars can naturally acquire very small masses of $\sim
$(TeV$)^2/M_{pl} \sim$(mm)$^{-1}$,
which is precisely in the interesting range for being tested by upcoming
sub-millimeter measurements of
gravitational
strength forces. The possibility that these experiments might uncover
``Yukawa moduli" has previously
been discussed in \cite{gian},  it is amusing that ``Yukawa
messengers" may also be probed as we will discuss more fully below.
Alternately, it could be that there is no SUSY in the bulk and that
$\chi$ starts massless but receives radiative corrections which do
not quite drag it back up to $M_s$ due to couplings or loop
factors; it is certainly easy to imagine that $\chi$ is light
within one or two orders of magnitude of $M_s$. In this case, the
suppressions in $\Delta_n(r)$ may either be exponential or purely
geometrical depending on whether the relevant $r$ is bigger or
smaller than $m_{\chi}^{-1}$.

Summarizing, in all cases, the smallness of the Yukawa coupling is
understood
from this very simple picture: the breakings of the flavor
symmetries are of $O(1)$  on far away branes, and the smaller
intensity of breaking on our branes is simply due either to the
geometrical power law fall off (for massless particles and $n>2$),
or the exponential decay characteristic of massive particles.

Up to now we have been assuming that the other wall is also a 3-brane.
We can however consider the possibility that it is an arbitrary
$p-$brane,
where $p \geq 3$. For $p<3$, different regions of our universe
will have different
transverse distances to the other brane, and the Yukawa couplings in our
universe would vary
unacceptably in different regions of the universe.  Of course, for $n$
``large" dimensions, the extra
$(p-3)$ dimensions of the $p$-brane are not infinite but have sizes
$\sim r_n$, and we must have $p \leq (n+3)$.
The only difference with the case of $p=3$ is that the $p-$brane no
longer
looks like a point in the $n$ transverse dimensions; rather, it
appears as a $p-3$ brane. The value of $\langle \chi \rangle$
``shined" by sources on the $p$-brane will be the potential set up
by a $p-3$ brane in the transverse $n$ dimensions, which by the
same symmetry arguments used above is the Yukawa potential set up in the
$n+3-p$ dimensions transverse to the $p-$brane. Therefore we must
replace
\beq
\Delta_n(r) \to \Delta_{n + 3 - p}(r)
\eeq
in the Yukawa coupling suppressions.
For $p>3$, this gives an enhancement of
the induced Yukawa coupling over the $p=3$ case, as expected.

Of course, it is also possible that the bulk messengers themselves
are not free to propagate in all the extra dimensions,
but only $n_{mess} \leq n$ of them. As long as they can propagate to the
other brane, they will still mediate flavor breaking to our wall,
and all the previous results hold replacing $n \to n_{mess}$.

It is clear that this mechanism can be generalized to explain the
small size of the ``spurion" $\epsilon \sim \phi/M_F$ of any flavor
model.
We just replace what was formerly considered the small vev of a
flavon $\phi$ relative to a higher scale $M_F$ by the
distance-suppressed
vev of a bulk messenger $\chi$ with identical $G_F$ quantum numbers as
$\phi$. Of course, the question of the origin of fermion mass hierarchy
is
transmuted in our picture to the question of what determines the
inter-brane separations. We do not have anything specific to say on this
dynamical question. Note however that in the case of
exponential suppressions, or even the power law suppressions
obtained for $n=6$,$\sim r^{-4}$, the walls do not need to more
than $O(10)$ times the fundamental Planck scale  away from each other to
cover
the observed range of fermion masses, so that at least
no large hierarchies in inter-brane separation are needed.
One can for instance imagine that there is a moderately large
extra dimensions $\sim 10$ times larger than the fundamental
scale, and that the different walls are stuck at different
orbifold fixed points in this extra dimension. Alternately, there could
be
dynamical mechanisms where attractive forces between the branes
form stable bound systems, much as the planets rotate in stable
orbits around the sun.

A different direction for getting small Yukawas
is to imagine different SM generations living on different walls.
In this scenario, the SM gauge fields must be delocalized
and free to propagate at least between the branes where the different
generations are trapped.
Again, massive modes linking the different
branes will give rise to exponentially small Yukawa couplings.
 Such possibilities have been noted within the context of orbifold
compactifications
of string theory.
This scenario seems to have
phenomenological difficulties in the scenario where  $M_s \sim $ TeV.
 The reason is that since the cutoff is $\sim 1$ TeV, if the exponential
suppression is to give small
 Yukawa couplings, we need to
have walls perhaps as far as $\sim (100)$ GeV$^{-1}$. Since the SM gauge
fields
must be delocalised in the extra dimensions on this scale, these
effects would have already been experimentally observed. Also, it
seems to not be as easy to avoid flavor-changing problems in this
scenario.
Of course, these problems can be solved by pushing the higher
dimensional Planck scale to
sufficiently  but it is clear that this scenario is
less safe than one where all the SM fields are localised to an at least
$M_s^{-1}$ thickness wall.

\section{Neutrino masses and SUSY breaking}
It is obvious that the mechanism for generating
small parameters presented above is generic and can be used to explain
many sorts of
small parameters.
For instance, we can generate
small Majorana
neutrino masses if lepton number is broken on far away walls.
This idea, together with other intrinsically higher dimensional
mechanisms for small neutrino masses will be discussed in
\cite{giajohn}.
One may perhaps hope to to be able to generate exponential
hierarchies of scales in this way, perhaps even for the weak-Planck
hierarchy.
It is certainly easy to generate a Higgs mass parameter exponentially
smaller than $M_{Pl}$ \cite{Turok},
the difficulty still is that in the absence of SUSY, radiative
corrections would make $\sim M_{Pl}^2$
contributions to the higgs mass$^2$. Therefore, some sort of SUSY may
still be needed.
Nevertheless, we can still construct very simple theories of
exponentially small SUSY breaking.
Indeed, the most trivial way of breaking SUSY is with a theory of a
single chiral field $\phi$ with a linear
superpotential $W= \lambda M_{pl}^2 \phi$.  We can generate an
exponentially
small $\lambda$ in the same way we
generated exponentially small Yukawas in the last section, by coupling
$\phi$ to a massive field in the bulk which
in turn couples to a (SUSY preserving) Planck-scale vev on a different
wall. We emphasize that
this mechanism is different than breaking SUSY on the other wall and
communicating the information
to our wall with bulk messengers \cite{Peskin}; here the dynamics on the
other wall preserves SUSY, but acts as a source for a massive bulk field
whose exponentially suppressed vev on our wall becomes a linear coupling
in the superpotential that
breaks SUSY directly on our wall.

\section{Constraints and signals from new light states}
\subsection{Distant familons}
So far we have ignored the dynamics on the other branes, other than
to assume that they provide an explicit source of $O(1)$ $G_F$
violation. If $G_F$ is a continuous global symmetry, however, the
walls also contain the goldstones of $G_F$ breaking, the
``familons" with decay constants $ \sim M_s$. Familons with low
decay constants are a disaster because they can be produced in
dangerous flavor-changing transitions such as $K \to
\pi +$ familon, which typically force the decay constants above $\sim
10^{12}$
GeV. One might hope that since in our case the familons live on
different branes, they have suppressed couplings and are therefore
harder to produce. However, at low energies, the distance between
the walls can not be resolved; the couplings of the familons to
the  SM fields are then dictated as usual by the non-linear
realization of $G_F$. More explicitly, if the source for $\chi$
on the other wall depends on massless fields as $\langle \phi \rangle
\to exp(i \pi^a T^a)
\langle \phi \rangle$, the ``shined" vev of $\chi$, and therefore the
Yukawa couplings on our wall, also
depends on these massless fields precisely as dictated by the
non-linear realization of the symmetry. Of course this is not a
problem if we wish to consider theories with high fundamental
scales $M_s \geq 10^{12}$ GeV, but for
theories with much lower $M_s$, this issue needs to be
addressed. One's first thought is to gauge the symmetry in order
to eat the familon. This does not work because the gauge boson
would then necessarily be a bulk field with very weak coupling to
wall states, only picking up an $\sim M_s^2/M_{Pl}$ mass from $\sim
M_s$
breaking on a three-brane \cite{ADD2}. For $M_s \sim$ TeV this is
so light that the familon (longitudinal component of the gauge boson)
can still be produced in decays. The simplest solution is to have
only discrete flavor symmetries, and therefore no light states
on the other walls. Note that in the usual theories of flavor,
simply decreeing that the true flavor symmetry is discrete does
not guarantee the absence of light familons, since the
renormalizable interactions of the theory typically admit
accidental continuous symmetries which are then spontaneously broken
yielding
pseudo-goldstone bosons. Higher dimension operators preserving the
discrete symmetry
but violating the accidental continuous symmetry will give mass to
the pseudo-goldstones, but these will be suppressed by a power of the
ratio of the flavor breaking scale to that
of the higher dimension operators. This need not be the case in
our scenario, since we are maximally breaking the flavor symmetry
at the fundamental scale $M_s$, so that the higher dimension operators
distinguishing the
continuous from discrete symmetries are unsuppressed.
Note that this effectively allows us to use continuous $G_F$: one
can always imagine that we are really considering an arbitrarily
large discrete subgroup of $G_F$, to avoid the problem with
familons on the other wall, while keeping the same predictions for
flavor-violations on our wall to arbitrarily high accuracy.
From the point of view of the low-energy theory,
there is only explicit violations of $G_F$ on the other walls, with no
other
light states to indicate spontaneous breaking.
\subsection{Light messengers of flavor breaking}
In the case where $\chi$ is very light, perhaps only getting a $\sim
(1$mm$)^{-1} \sim 10^{-3}$ eV mass,
we have the interesting possibility of observable sub-millimeter forces
mediated by $\chi$ exchange.
Recall that the couplings of $\chi$
to SM fields is of the form
\beq
{\cal L}  \sim \int d^4 x f_i f^c_{j^c} H(x)
\frac{\chi_{i j^c}(x,y=0)}{M_s^{(n+2)/2}}
\label{chic}
\eeq
Expanding
\beq
\chi(x,y=0) = \langle \chi \rangle (x,y=0) + \chi'(x,y=0)
\eeq
it is readily apparent
$\chi'$ can in general have both flavor-changing and flavor-violating
couplings.
For $n>2$, all the KK excitations of $\chi'$ are too heavy to be
relevant to sub-millimeter force experiments, so
only the coupling of the zero mode is needed. Since $\chi'$ is a bulk
field, this coupling is suppressed by
the value of the wavefunction of
the zero mode on the wall $\sim 1/\sqrt{V_n}$.  More formally,  $\chi'$
is Fourier expanded as
\beq
\chi'(x,y=0) = \sum_{k_1,\cdots,k_n} \frac{1}{\sqrt{V_n}}
\chi^{k_1,\cdots,k_n}(x)
\eeq
The long range force that is generated from the Yukawa coupling to the
zero mode
\beq
\rho f f^c \chi^{0}, \, \rho= \frac{v}{M_{pl}} \sim 10^{-16}
\eeq
where we have used the relation $M_{pl}^2 \sim M_s^{n+2} V_n$.
Note that this Yukawa coupling is both $n$ and $M_s$ independent.
While $\rho$ seems miniscule, it dominates gravity by a factor of $\sim
10^6$ at distances
shorter than the $1/m_{\chi}$. For example,  the ratio of $\chi$
exchange force to gravity for nucleons is
\beq
F_{\chi}:F_{\mbox{grav}} = \rho^2:G_N m^2_{\mbox{nucleon}} \sim 10^6
\eeq
In this respect the $\chi'$ field mediates a force of the same order of
magnitude as gauge bosons in the bulk coupled to
a linear combination of $B,L$ as discussed in \cite{ADD2} and more fully
discussed in Section 7.
There is one difference: the effective 4-dimensional gauge coupling is
$g \sim M_s/M_{pl}$, and
grows as $M_s$ is pushed above a TeV. On the other hand, as remarked
above, the strength of the
Yukawa force is independent of $M_s$. This spectacular signature of
sub-mm forces a million times stronger than gravity
only depends on having $\chi's$ sufficiently light that their Compton
wavelength falls in the range
$\sim 1 \mu m - 1 mm$ soon to be probed by experiment.
Furthermore, the force mediated by $\chi$ exchange can be distinguished
from
similar size bulk gauge field induced forces, since the latter is
repulsive while the
former is attractive .

It is instructive to compare this ``Yukawa messenger" force with the
``Yukawa modulus" force discussed in
\cite{gian}.
In \cite{gian}, supersymmetric theories were considered with a
continuous global flavor symmetry
$G_F$ spontaneously broken at the usual 4-d Planck scale. The ``Yukawa
moduli"
are just the goldstones of $G_F$ breaking. In a supersymmetric theory,
there is a full goldstone chiral multiplet $\Phi^a$ for each broken
generator $X^a$ of $G_F$.
The complex scalar $\phi^a = \sigma^a + i \pi^a$ includes the usual
goldstone field ($\pi^a$) together with a scalar partner
$\sigma^a$ whose mass is only protected by unbroken SUSY.  If SUSY is
broken at very low energies, these fields can pick up
a small mass $\sim$(TeV$)^2/M_{pl} \sim (1$mm$)^{-1}$ and can mediate
interesting sub-mm forces.
The linear couplings of $\phi^a$ are fixed by the non-linear realisation
of $G_F$ to be given by
\beq
W \supset \phi^a \delta_{X^a} (\lambda^{i j^c} f_{i} f^c_{j^c})
\eeq
where $\delta_{X^a}(f)$ denotes the first order variation of $f$with
respect to the
broken generators $X^a$ of $G_F$.
The crucial point is that the couplings of $\phi^a$ are suppressed by SM
particle masses.
Of course this is because all couplings must vanish in the limit where
the SM chiral symmetries are unbroken.
The force mediated
by the scalar partners of the goldstones are therefore truly
gravitational in strength. By contrast, the force mediated by the
``Yukawa messengers" are present even if $\chi$ does not acquire a vev
on our wall, and the strength of the force is enhanced
by the $\sim 10^6$ factor we found in the previous paragraph.

Given the obvious interest in these signals, and their
crucial reliance on a light $\sim ($mm$)^{-1}$ mass,
it is important to insure that such light $\chi$'s are
not excluded on other grounds.
As we have remarked, the sub-mm force is actually independent of
the value of the fundamental scale $M_s$, while the dangerous production
of
bulk $\chi$ modes is suppressed by powers of $M_s$. Therefore, we can
always take
$M_s$ high enough to avoid phenomenological problems while keeping the
sub-mm signal.
What we are interested here is how low $M_s$ can be, and in particular
whether
$M_s \sim $ TeV is allowed.
Indeed, we will find quite significant constraints
even on the flavor conserving interactions of the $\chi$ fields from
astrophysics and cosmology, while the flavor-violating couplings
are even more severely from rare
decays. Nevertheless, $M_s > 10^6$ TeV satisfies all constraints in all
cases,
while $M_s \sim $ TeV is allowed for the cases $n \geq 4$.

Let us begin with the flavor-conserving $\chi$ interactions, considering
the bounds from astrophysics and cosmology which result from the
overproduction
of bulk $\chi$ modes. Let us first perform the analogue of the
dimensional
analysis done in \cite{ADD2} to determine the scaling of
the rate for $\chi$ production with the temperature $T$ of
wall states.
From the couplings given in
eqn.(\ref{chic})
this rate is determined by dimensional analysis to be
\beq
\mbox{rate for $\chi$ prod.} \sim \frac{v^2 T^{n-2}}{M_s^{n+2}}
\eeq
where $v \sim 175$ GeV is the Higgs vev.
Notice that just like gravitons, this rate grows softer in the infrared
as $n$ increases, as is to be expected for a bulk mode.
However, it is not quite as soft as the rate for graviton production
\beq
\mbox{rate for grav. prod.} \sim \frac{T^n}{M_s^{n+2}}
\eeq
Therefore, the constraints on this scenario are stronger than the
corresponding
ones for gravitons. In fact, for $M_s \sim$ TeV, we can
roughly see that the constraints on $\chi$ for $n$ extra dimensions are
the
same as for gravitons with $n-2$ extra dimensions.
For gravitons, the constraints for $n=2$ from the Supernova
forces $M_s \sim 30$ TeV, while $M_s \sim$ TeV was safe for $n>2$.
Therefore, the constraints from $\chi$ overproduction in the SN
rule out the cases $n=2,3$, force $M_s \geq 30$ TeV for $n=4$ and
can have $M_s \sim$ TeV for $n \geq 5$.

Similarly, the bounds on the "normalcy" temperature $T_*$ in the early
universe coming
from $\chi$ evaporation, overcooling the universe, are the same as
gravitons
with $n$ replaced with $n-2$, once again ruling out $n=2,3$, while
$n\geq 4$
has $T_* \geq 10$ MeV, leaving nucleosynthesis safely unaltered.
The strongest cosmological constraint on gravitons arose because
their lifetime for decaying back into wall states exceeded the age of
the
universe, so that overclosure and the distortion of the
background gamma ray spectrum from their decays had to be considered.
In our case, however, these bounds do not apply. The reason
is that, since the $\chi$ fields are more strongly coupled, they decay
more quickly. Indeed, the width for a given KK excitation
of $\chi$, produced at temperature $T$ (and therefore
having a mass from the 4-d point of view $\sim T$)
to decay to SM states on the wall is
\beq
\Gamma \sim \frac{v^2}{M_{pl}^2} T
\eeq
which gives a lifetime for $\chi$ ranging between 10 to 10,000 years for
$T \sim 1$ GeV - 1 MeV. Since these times are before recombination,
the decay products are harmless and rethermalise. There is also no
worry about the decay products destroying weakly bound states like
deuterium.
Their lifetime is long enough that, by the time they decay,
the universe is so dilute that the decay products can not interact with
enough deuterons to
destroy significant numbers of them.

Summarizing, the constraints on the flavor-conserving $\chi$
interactions
rule out $n=2,3$ (at least for $M_s \sim $ TeV), while they force
$M_s \sim 30$ TeV for $n=4$ and are fine for $M_s \sim$ TeV for $n \geq
4$.

Far more important constraints result from the flavor-violating
couplings of
$\chi$.  Most worrisome are dangerous processes like $K \to
\pi + \chi_{sd}'$
or $\mu \to e + \chi_{\mu e}'$.
Consider the decay $K \to \pi + \chi'$. By the dimensional
analysis familiar from \cite{ADD2}, we have
\beq
\Gamma(K \to \pi + \chi') \sim \frac{v^2 m_K^{n+1}}{M_s^{n+2}}
\eeq
where $v \sim 250$ GeV is the Higgs vev. Requiring that the
branching ratio be less than the experimental limit of $\sim 7 \times
10^{-9}$
puts a lower bound on $M_s$:
\beq
M_s > 10^{\frac{31 - 3n}{n+2}} \mbox{TeV}
\eeq
which ranges from $\sim 10^{6}$ TeV for $n=3$ to $\sim 30$ TeV for
$n=6$. We see that if we wish to have $M_s \sim$ TeV, the mass of the
$\chi$ field coupling to strange and down squarks must
be pushed above $\sim 1$ GeV in order not to be produced in
Kaon
decays. The branching ratio for $\mu \to e +$ familon is $< 1 \times
10^{-10}$, and
an identical computation for $\mu \to e + \chi_{\mu e}'$ yields
the bound
\beq
M_s > 10^{\frac{26 - 4n}{n+2}} \mbox{TeV}
\eeq
which ranged from $\sim 3000$ TeV for $n=3$ to $\sim 2$ TeV for
$n=6$. It is more conceivable that such a light $\chi$ could
couple to the lepton sector for TeV scale $M_s$ in this case, and
sub-millimeter ``Yukawa messenger" signals in the lepton sector
are still consistent with $M_s \sim $ TeV.

\section{Novel Higgs decays}
All the constraints from rare decays, astrophysics and cosmology
disappear if
the $\chi$ fields are heavier than a few GeV. On the other hand, if they
are this heavy, they
will not give rise to signals for
the upcoming sub-millimeter force experiments. On the other hand, if the
$\chi$'s are lighter than any physical
Higgs modes, they can be produced in novel Higgs decays with significant
rates.  It is normally
believed that the Higgs couples most strongly to the heaviest
generations, since they have the largest Yukawa
coupling. Therefore, the dominant decay mode for the neutral Higgs with
mass $m_{H^0} < 2 m_t$ is to $b \bar{b}$, with
width
\beq
\Gamma(H^0 \to b \bar{b}) = \frac{\lambda_b^2}{16 \pi} m_H
\eeq
However, in our scenario, the Yukawa couplings are the vevs of $\chi$ on
our wall, and expanding as usual
$\chi(x,y=0) = \langle \chi \rangle(x,y=0) + \chi'(x,y=0)$, the relevant
interactions of $H$ with the SM fermions and
$\chi'$ are given by
\beq
{\cal L} = \int d^4 x \lambda_{i j^c} f^{i} f^{j^c} H + \kappa
\frac{\chi'_{i j^c}}{M_s^{(n+2)/2}}(x,y=0) f^i f^{j^c} H
\eeq
where we have restored the dependence on the fundamental scale $M_s$ and
$\kappa$ is a dimensionless coupling,
$\lambda_{i j^c} = \kappa \langle \chi_{i j^c} \rangle$. We see that
while the coupling of $H$to $f^i f^{j^c}$
alone is suppressed by the Yukawa coupling for lighter generations, the
couplings of $H$ to $f f^c \chi'$ are
are not suppressed by small Yukawas but only by the scale $M_s$. The
width for $H \to f f^c \chi$, where
$\chi$ escapes into the bulk as missing energy, is easily estimated
\beq
\Gamma(H^0 \to f f^c \chi) \sim \kappa^2 \frac{m_H^{n+3}}{16 \pi
M_s^{n+2}}.
\eeq
This gives a branching fraction relative to the usual $b \bar{b}$ mode
of
\beq
\frac{\Gamma(H^0 \to f f^c \chi)}{\Gamma(H^0 \to b \bar{b})} \sim
\frac{\kappa^2}{\lambda_b^2}
\left(\frac{m_H}{M_s}\right)^{n+2}.
\eeq
For relatively low $n=2,3$, this mode can have a significant branching
fraction for $M_s/\kappa \sim 1$ TeV. For instance,
we can have $O(1)$ branching fractions for $M_s/\kappa \sim 1$ TeV for
$n=2,m_H \sim 100$ GeV or $n=3, m_H \sim 200$ GeV. Furthermore, the
decays
$H^0 \to e (\mu)^+ e(\mu)^- + (\chi=$missing energy) are very clean in a
hadronic environment, where the
$b \bar{b}$ signal is not. Indeed, recall that the usual signal for
Higgs production at Hadron colliders is
via $H^0 \to \gamma \gamma$ which typically has a tiny branching
fraction $\sim 10^{-3} \to 10^{-4}$.
These new Higgs signals have enormous branching fractions in comparison
and may be the dominant discovery mode for the Higgs at hadron
colliders, although a more detailed analysis
is clearly needed. Since the branching ratio for decaying to bulk modes
increases with the mass of the decaying
particles, even larger widths are possible for the charged Higgs modes
in two Higgs doublet models, for instance $H^+ \to e^+$ or $\mu^+$ + $\chi$.
Of course the usual decay to top+bottom provides a bigger SM background.
The $f f^c H \chi$ coupling also gives rise to new Higgs production mechanism.
In any, this sort of novel Higgs physics, especially in the decay to light 
generation fermions + missing energy, is certainly a smoking gun for our
scenario for generating flavor at TeV
energies.

\section{Avoiding large FCNC problems}

If all possible higher dimension operators suppressed by $\sim 1$ TeV
are present in the SM, there are not only disastrous problems with
proton decay
but also with flavor-changing effects.
The most challenging flavor-changing constraints arise from the Kaon
system.
If we consider 4-fermion $\Delta S = 2$ operators of the form
$C$/(1 TeV$)^2 {\cal O}$, the bound
on the coefficient $Re C$ from $\Delta m_K$ are \cite{ACHM}
\begin{eqnarray}
{\cal O} = (\bar{d}^A \bar{\sigma}^{\mu} s_A)^2&,&Re C<4 \times 10^{-7}
\nonumber
\\
{\cal O} = (d^{cA} s_A)^2 &,&Re C<6 \times 10^{-8} \nonumber \\
{\cal O} = (d^{cA} s_B)(d^{cB} s_A) &,& Re C<3 \times 10^{-7} \nonumber
\\
{\cal O} = (d^{cA} s_A)(\bar{d}^B \bar{s^c}_B) &,&Re C<5 \times 10^{-8}
\nonumber
\\
{\cal O} = (d^{cA} s_B)(\bar{d}^B \bar{s^c}_A) &,&Re C<2 \times 10^{-7}
\label{ops}
\end{eqnarray}
where $A,B$ are color indices.
The constraints on $Im C$ from $\epsilon_K$ are $\sim 100$ times
stronger.
Any theory of new physics at the weak
weak scale must explain why the coefficients of these operators are so
small. Indeed, this is as much a part of the flavor problem as
explaining the
smallnes of the dimension-4 Yukawa couplings.
In this section, we will explore this issue in the context of our
mechanism for
generating the fermion mass hierarchy.

Before proceeding, let us remember the origin of the problem.
In the limit where all Yukawa couplings are set to zero, the SM gauge 
interactions admit a $U(3)^5$ flavor symmetry rotating the three
generations
of $Q,U^c,D^c,L,E^c$. The three Yukawa couplings matrices
$\lambda_{U,D,E}$
explicitly break this $U(3)^5$  and are the ${\it only}$ sources
of breaking within the SM. This is reason for the successfully predicted
small flavor violation in the SM,
the heart of the GIM mechanism.
Consider for instance operators of the
form
\begin{equation}
\frac{1}{16 \pi^2 m_W^2}
\bar{Q}^a \bar{\sigma}^{\mu} Q_b \bar{Q}^c \bar{\sigma}_\mu Q_d C^{ac}_{bd}
\end{equation}
where $a,\cdots,d$ are $U(3)_Q$ flavor indices.
This is the structure of the operator generated in the usual box
diagram.
If the only sources of
flavor-violation are in the Yukawa matrices $\lambda_{U,D,E}$, purely
on gounds of $U(3)^5$ transformation properties we must have that
\begin{equation}
C^{ac}_{bd} =  \delta^a_c \delta^b_d +
(\lambda_{U,D}^{\dagger} \lambda_{U,D})^{a}_b \delta^c_d +
(\lambda_{U,D}^{\dagger} \lambda_{U,D})^{a}_b
(\lambda_{U,D}^{\dagger} \lambda_{U,D})^{c}_d + \cdots
\end{equation}
where we have omitted overall coefficients in front of each term,
and other symmetric terms.
For the $\Delta S=2$ operator, we want $C^{12}_{12}$ in the basis where
$\lambda_D$ is diagonal. The only non-zero contribution
arises from
\begin{equation}
C^{12}_{12} \sim [( \lambda^{\dagger}_U \lambda_U)^{1}_2]^2 \sim
\lambda_c^4 \theta_c^2.
\end{equation}
This does not look like the usual GIM suppression factor of $\Delta m_K$
which is $\sim (m_c/m_W)^2$. However, we have to remember that in the
usual
box diagram, there are light particles running in the loops, and that in
fact
there is an infrared divergence cutoff by the charm mass
which gives an enhancement $\sim 1/\lambda_c^2$.
If physics beyond the SM at the TeV scale has no
new sources of flavor violation beyond
$\lambda_{U,D,E}$, then there is not even any infrared
enhancement of this operator, which then makes negligibly
small contributions to $\Delta m_K, \epsilon_K$.
A similar operator analysis can be made to
estimate the coefficients of all the other operators in
eqn.(\ref{ops}), with the same conclusion. We have learned that
as long as the only flavor violation is that given by the
Yukawa matrices of the SM, there are no flavor-changing
problems with (TeV$)^{-1}$ suppressed operators.
This is a well-known fact to flavor model builders (see e.g.
\cite{HallRand}).
The trouble is that
generically, extensions of the SM do have $U(3)^5$ violating operators
beyond
the $\lambda'$s. For instance, in SUSY, we have
the 5 scalar mass matrices $m^2_{Q,U,D,L,E}$.
New invariants in principle unrelated to the Yukawas
appear in the coefficients of flavor-violating
operators.
For instance, the usual squark box diagrams
give
\begin{equation}
C^{12}_{12} \sim \left(\frac{(m^2_Q)^{1}_{2}}{4 \pi m_{SUSY}^2}\right)^2
\end{equation}
with a strong resulting bound on the off-diagonal elements of the squark
mass
matrices $\sim 10^{-2} -10^{-3}$ depending on the operator.
Of course, one could take the attitude that $U(3)^5$ is a flavor
symmetry
broken only by fields whose vevs are the Yukawa matrices
\cite{HallRand}. This resolves the
flavor-changing problem, but leaves the origin of the hierchies in the
Yukawa themselves unexplained. In a four-dimenisonal theory, this
question
would be relegated to the unknown dynamics of flavor symmetry breaking.
As we will see below, in our picture,
we can use a $U(3)^5$ flavor symmetry to control flavor-changing, while
simultaneously offering an understanding of the hierarchies as
coming from distant flavor breaking.

We take the flavor symmetry to be $G_F=U(3)^5$ \footnote{Of course, we
really
mean a sufficiently large discrete subgroup of $U(3)^5$, see
Section 3.1.}. The idea is to
have a different wall generate each non-zero element of the
Yukawa matrices $\lambda_{(U,D,E) i j^c}$.
We can then label the walls by the indices $A k l^c$, (where $A=U,D,E$),
a
distance $r_{A k l^c}$ from our wall.
The hypothesis is that on the $(A k l^c)$'th wall, there is a
wall-localized
field $\phi^{(A k l^c)}_{i j^c}$, with the $i, j^c$ indices transforming
under $G_F$ like the Yukawa matrix $\lambda_A$.
These fields are assumed to have an $\sim M_s$ vev of the form
\beq
\langle \phi^{(A k l^c)}_{i j^c} \rangle
\sim M_{s} \delta^k_i \delta^{l^c}_{j^c}
\label{vevs}
\eeq
The bulk messengers are $\chi_{A i j^c}$ which also transform as
$\lambda_A$
under $G_F$.
The interactions responsible for transmitting flavor-breaking
to our wall are
\begin{eqnarray}
{\cal L} &\supset& \int_{\mbox{us}} d^4 x Q^i H U^{cj}(x) \chi_{U i
j^c}(x,y=0) + \mbox{sim. for D,E} \nonumber \\
&+& \sum_{A k l^c} \int_{A k l^c \mbox{wall}} d^4 x_{Akl^c}
\phi^{Akl^c}_{ij^c} \chi^{* A i j^c}(x_{Akl^c},y=y_{Akl^c})
\end{eqnarray}

As usual, the sources $\langle \phi \rangle$ will set up a classical
profile for $\chi_{A i j^c}$,
\beq
\langle \chi_{A i j^c} \rangle (x,y) = \sum_{k l^c} \langle \phi^{A k
l^c}_{
i j^c} \rangle \Delta_n(y - y_{A k l^c}) \sim \Delta_n(y - y_{A i j^c})
\eeq
where we are once again using units with $M_s = 1$ and we used the
specific
form of the vevs in eqn.(\ref{vevs}) to get the second equality.
The value of $\langle \chi \rangle$ on our wall determines the Yukawa
matrices:
\beq
\lambda_{A i j^c} = \langle \chi_{A i j^c} \rangle(x,y=0) \sim
\Delta_n(r_{A i j^c}).
\eeq
For massive $\chi$'s, the inter-wall distances can be within $O(10)$ of
the
fundamental scale while covering the range of observed Yukawa couplings.

We now wish to show that despite the fact that $U(3)^5$ is maximally
broken
on far away walls, the only flavor violation felt on our wall are given
by the "shined" Yukawa matrices $\lambda_A$. This is perhaps surprising,
since
the classical profile of $\chi$ certainly knows about flavor breaking 
throughout the bulk between our wall and the others where flavor is
broken at
$O(1)$. However, if the propagation of $\chi$ in the bulk is linear,
 i.e. if $\chi$ propagates as a free field in the bulk, it is obvious
that
the {\it only} sources of flavor violation in the theory are the vevs
$\langle \chi_{A} \rangle (x,
y=0) = \lambda_A$. Explicitly, the Lagrangian is of the form
\begin{eqnarray}
{\cal L} &=& \int_{\mbox{us}} d^4 x Q^i U^{c j^c} H \chi_{Q i
j^c}(x,y=0)
 + \cdots + \mbox{higher dim. ops.} \nonumber \\
&+& \sum_{A i j^c} \int d^4 x_{A i j^c} \mbox{Tr} \langle
\chi^{\dagger A}(x,y_{A i j^c}) \phi_A \rangle \nonumber \\
&+& \int d^4 x d^n y \mbox{Tr} \left( \partial \chi^{\dagger A}
\partial \chi_A -
m^2_{\chi_A} \chi^{\dagger A} \chi_A \right).
\end{eqnarray}
If we now expand $\chi=\langle \chi \rangle + \chi'$, we find
\begin{eqnarray}
{\cal L} &=& \int_{\mbox{us}} d^4x Q^i U^{c j^c} H (\lambda_{Q i j^c}
 + \chi'_{Q i j^c}(x,y=0)) + \cdots \nonumber \\
&+& \int d^4x d^n y \mbox{Tr} \left(\partial \chi^{\prime \dagger A}
\partial \chi'_A - m^2_{\chi_A} \chi^{\prime \dagger} \chi'_A \right)
\end{eqnarray}
The important point is that all the flavor-violating interactions in
$\langle
\chi \rangle(x,y)$ at points away from $y=0$ have disappeared.
All of the higher dimension operators involving SM fields and $\chi$ on
our wall
then become $U(3)^5$ invariants with $\lambda_A$ acting as spurions,
and we are therefore manifestly safe from FCNC worries induced by the
physics
above $M_s$ even for $M_s \sim$ TeV.

This situation changes somewhat if the $\chi'$s have self-interactions
in the
bulk. As an example, suppose the Lagrangian contains a term of the form
\beq
\int d^4x d^n y h \mbox{Tr} (\chi^{\dagger} \chi)^2
\eeq
After expanding $\chi = \langle \chi \rangle + \chi'$, we have
interactions of the form
\beq
\int d^4x d^n y h \mbox{Tr} \left(\chi^{\prime \dagger} \chi'
\chi^{\prime \dagger}
\langle \chi \rangle \right)
+ \cdots
\eeq
which are now clearly sensitive to the profile of $\langle \chi \rangle$
away
from $y=0$. The SM fields can then "sniff" breakings of flavor other
than the
Yukawas through emitting $\chi$'s into the bulk which feel $\langle \chi
\rangle$ away from $y=0$ through non-linear interactions. These are not
uncontrallably
large effects, however. In order to "sniff" a value for $\langle
\chi(x,y)
\rangle$ significantly different from $\lambda$ which is its value at
$y=0$, we must propagate far into the bulk and use the $h$ interaction.
However, the propagators to get to this far away point are themselves
suppressed.It is a detailed model-builidng question, having to do with
the precise
configuration of the various walls, whether these effects are harmful.
We will not investigate this issue here however, as it is clearly a
higher
order question. Reiterating: if the $\chi's$ have no self-interactions
in the bulk,
there is no problem, while the problem is not uncontrollably
re-introduced even
if there are $O(1)$ self-interactions.

It is evident that more elegant theories of flavor, requiring fewer
fields and
distant walls, can be constructed. A new and more restrictive framework
for
flavor model building is suggested in our framework.
Given any $G_F$ and breaking pattern parametrized by spurions, the small
size of the spurions can always be explained by
breaking flavor on distant walls.
The challenge is to formulate a theory where, remaining fully agnostic
of
TeV scale physics and therefore allowing all $G_F$
invariant operators suppressed by (TeV$)^{-1}$, flavor changing effects
are
small enough.

This requirement is stronger than,say, requiring flavor
symmetries to solve the SUSY flavor problem. For instance, it is known
that
a $U(2)$ flavor symmetry acting on all SM fields of the first two
generations
can solve the SUSY flavor problem by guaranteeing sufficient squark
degeneracy
\cite{U2}. But $U(2)$ fails to satisfy the new criteria. Even before
breaking, the following $U(2)$ invariant operator (a,b are $U(2)$
indices)
\beq
(Q_a D^c_b)(\bar{Q}^b \bar{D}^{ca})
\label{bad}
\eeq
contains a $\Delta S=2$ operator taking e.g. $a=2,b=1$.
This therefore requires $M_s > 1000$ TeV to be safe. Why is this
operator
not generated in the SUSY theory? The reason is that in the limit where
$U(2)$ is unbroken, all Yukawa couplings involving the first two
generations
vanish. The renormalizable interactions of the theory then have an
enlarged
$U(2)^5$ symmetry which forbids $\Delta S=2$
transitions. Operators like the one in eqn.(\ref{bad}), which are are
non-zero even before
$U(2)$ breaking, can therefore never be generated. With SUSY,
we are not dealing with an arbitrary theory at TeV energies.
Rather, we have a perturbative, renormalizable QFT, which does not
generate
all operators consistent with low-energy symmetries. In our case,
however, the
theory at the TeV scale is the short distance theory of gravity and is
unknown.
While it is also possible that in a specific theory not all operators
are generated, we can not assume this in any controlled way. We must
therefore
insist that the group theory and breaking pattern alone
suppress dangerous flavor-changing effects. A good starting point may be
a $U(2)_L \times U(2)_R$
symmetry, where the $U(2)_L$ acts on the electroweak doublets of the
first two generations while the $U(2)_R$ acts
on the remaining first two generation fields. This retains many of
the desirable features of $U(2)$ while
forbidding the dangerous $U(2)$ invariant operator we found above.
More investigation is however clearly necessary to assess the viability
of this model.
A theory based on the $S_3^3$ flavor symmetry is also worth considering
\cite{S3}.

\section{Gauged Symmetries in the bulk}
In \cite{ADD2}, the possibility was raised that baryon number is gauged
in the bulk.
It was shown that if broken only on a wall, the $U(1)_B$ gauge boson
could naturally get a $\sim$(mm)$^{-1}$ mass,
and could mediate repulsive forces $\sim 10^{6} - 10^{8}$ times gravity
at sub-mm distances.
It was also stated that if $U(1)_B$ is broken
on a different brane,
the proton decay rate can be enormously suppressed.
In this section we wish to explore this issue in more detail, for the
general case of some
arbitrary new gauge symmetry $U(1)_X$. Namely, if
$U(1)_X$ is broken on a different brane by the vev of some field $\phi$
charged under $X$,
what level of $X$-violation is generated on our brane?

As we have previously argued, in order to communicate the
information of symmetry breaking from one wall to the other, some
bulk field must be charged under $U(1)_X$.
More formally, consider local operaotrs $O_{our wall}(x)$ of charge $+q$
and $O_{other wall}(x')$ of charge $-q$.
After integrating out all bulk
modes, the only gauge invariant operator that can be induced must be of
the form
\begin{equation}
O_{our wall}(x) P e^{i q \int_{x,y=0}^{x',y=y_*}  A dl} O_{other
wall}(x')
\end{equation}
where the path from $(x,y=0)$ to $(x',y=y_*)$ is unspecified in the path
ordered exponential.
Since this operator by neccessity involves the gauge field in the bulk,
the gauge field lines must be able to end
in the bulk, i.e. there must be a charged field in the bulk.

Therefore, since the $U(1)_X$ gauge boson is neutral under $X$, if the
only bulk fields are
gravity and $U(1)_X$, even after the breaking
of $U(1)_X$ on the other wall, no $X$ violating operators are induced on
our wall.
The only possibility to induce $X$ violation is if there are charged
modes in the
bulk. As we have seen, if these modes are massive $\sim M_s$,
the coefficient of $X$ violating operators on the wall is
exponentially suppressed by $\sim e^{-M_s r_n}$.
If the wall where $U(1)_X$ is broken is maximally far away from ours
$i.e. r \sim r_n$, and if the heavy modes
 are Planck-scale in mass,  the suppression
can completely kill proton decay. For $U(1)_B$, the relevant proton
decay operator would be
\begin{eqnarray}
{\cal L}_{p decay} &\sim& e^{-(M_s r_n)} \frac{Q^{\dagger} U^c D^c
L^{\dagger}}{M_s^2} \nonumber \\
&\sim& e^{-10^{32/n}}  \frac{Q^{\dagger} U^c D^c L^{\dagger}}{M_s^2}
\end{eqnarray}
giving a lifetimg
\begin{equation}
\tau_p \sim 10^{10^{30/n}} \times \mbox{any units}
\end{equation}

If there is an additional light messenger
field in the bulk carrying $U(1)_B$,
there are two possible worries.
Of course, if this field is lighter than the proton, the proton
can decay in a B conserving way with an unsuppressed rate, and
so this must be forbidden. Even if the mass of this field is
pushed above the proton mass, the exponential suppression of the
induced B violating operator may not be sufficient.
Indeed, if we ignore the exponential suppression and just use the power
suppression
for $n>2$, the scale of suppression
for the p-decay operator  becomes
\begin{equation}
{\cal L}_{p decay} \sim \frac{1}{\Lambda^2} Q^{\dagger} U^c D^c
L^{\dagger}; \, \Lambda = M_{Pl}/(M_s r_n)
\end{equation}
since for a low string scale we need $M_s r_n$ much larger than 1,
the proton would still decay too quickly. Therefore, bulk scalars
with unit baryon number must be pushed significantly above the
proton mass.

For completeness, we now discuss the possibility that the gauge group
$G$ in the bulk is non-abelian.
Since the gauge bosons {\it are} charged under the gauge group, if $G$
is broken on a different wall,
some information of $G$ breaking can be transmitted to our wall. The
sort of information is however severely
constrained again by gauge invariance. The point is that any operator
induced on our wall by integrating out
the bulk gauge bosons must begin as a gauge invariant operator of the
form
\begin{equation}
O(\phi, D_{\mu} \phi) f(F_{\mu \nu}^a).
\end{equation}
When the gauge field lines eminating from this vertex travel to the
other wall and feel the breaking of $G$, this can turn into
into a $G$ breaking operator $O(\phi,\partial_{\mu} \phi)$ on the wall.
But note that $O$ can not transform
under an arbitrary non-trivial represenation of $G$: it is constrained
by the requirement that the above operator be gauge
invariant.  Only those operators can be generated which contain a
singlet in
the product with any number of
adjoints (contained in $f(F_{\mu \nu}^a)$).

Finally, we note that the bounds from overproduction of light gauge
bosons
in the bulk in astrophysical systems and in the early universe are
similiar to those discussed above for the flavor-conserving interactions
of
the light $\chi$ fields. For $M_s \sim $ TeV, they exclude the cases
$n=2,3$;
for $n=4$ they require $M_s > 30$ TeV, and are safe even with $M_s \sim
1$
TeV for $n \geq 5$.

\section{Accelerator signals for bulk scalars and vectors}
In the previous sections, we have motivated reasons for the
existence of scalar and vector fields in the bulk with
couplings to wall fields suppressed by the fundamental scale $M_s$: the
scalars could be messengers of
flavor breaking from distant walls, while the vectors could play a
role in stabilizing the proton. For $M_s \sim 1$ TeV and  the case where
these fields
are very light, there are significant constraints from
rare decays, astrophysics and cosmology which force minimally $n>3$.
However, if they  have masses
above a few GeV all these constraints disappear. Even in this
case, however, they can still be produced at colliders (where their
small mass is irrelevant
compared to the beam energy), much like bulk graviton production
\cite{AADD,gravbulk}.
Of course the event must contain a  photon or a jet to be visible,
so the sort of signal we are interested in are, in complete analogy with
the graviton case
\begin{eqnarray}
e^+ e^- \to \gamma + (\mbox{bulk vector or scalar = missing
energy}) \nonumber \\
q \bar{q} \to \mbox{gluon} + (\mbox{bulk vector or scalar = missing
energy})
\end{eqnarray}
In fact, just as the long-range forces mediated by these fields are
$\sim 10^6$
times stronger than gravity, and the astrophysical and cosmological
constraints on these fields
were stronger than the gravity case, their production cross section at
colliders can be comparable to or dominate the graviton production
cross-section.
Recall that the graviton cross sections at energies $E$ beneath $M_s$
scale as
\beq
\sigma(\mbox{grav. prod.}) \sim \frac{E^{n}}{M_s^{n+2}}
\eeq
whereas the scalar and vector cross sections vary as
\begin{eqnarray}
\sigma(\mbox{scalar. prod.}) \sim \frac{v^2 E^{n-2}}{M_s^{n+2}}
\nonumber \\
\sigma(\mbox{vector. prod.}) \sim \frac{E^{n-2}}{M_s^{n}}
\end{eqnarray}
Therefore, at energies beneath $v \sim 200$ GeV, the scalar
production is at least comparable to graviton production, while
vector production can dominate for all $E<M_s$. Very recent
complete analyses for the graviton production at present and
future colliders have been carried out in \cite{gravbulk} with
quite strong results; a similar analysis for bulk scalar and
vector production, while more model-dependent, will likely yield a
still more powerful probe of $M_s$ extending to larger values of $n$.

\section{Electroweak breaking with no Higgs}
In this section we wish to explore another novel possibility raised by
the 3-brane universe scenario.
Unlike many of the previous observations about flavor,
which operate for any value of the fundamental scale
$M_s$,
the considerations of this section,
being intimately related to electroweak symmetry breaking, require
$M_s \sim$ TeV.
Up to now, we have rather loosely been referring to the theory on the
wall as the
``Standard Model", with the tacit assumption of the presence of a light
higgs with a negative mass$^2$
driving electroweak symmetry breaking.
However, this need not be the case. Indeed, as is well known, all the
accurately tested aspects
SM phenomenology are reproduce by the weakly
gauged electroweak chiral Lagrangian. Unitarity breaks down in this
theory at energies $\sim 1$ TeV, and new physics must enter to unitarize
it at these
energies. But there is no reason for a light Higgs to unitarize the
theory. Indeed, in our scenario where strong gravity is brought down to
the TeV
scale, it is tempting to consider the possibility that physics related
to strong gravitational
physics at the TeV scale could trigger dynamical electroweak
symmetry breaking.
This is not implausible in
string theory
since it is likely that the theory is neither at weak coupling (since
the dilaton runs away), nor at strong coupling (since this is dual to
another weakly
coupled theory), but at intermediate strong coupling. It is not
unreasonable to expect that the theory at intermediate strong coupling
may show
qualitatively different behavior than that expected from perturbation
theory, perhaps including the formation of resonances and dynamical
symmetry
breaking. We will however
explore these ideas within a specific field-theoretic example,
``technicolor in the bulk". We will show that
even an abelian technicolor group in the bulk can naturally force TeV
condensates for technifermions localised on the wall,
triggering electroweak breaking.
We can then use ideas from the previous sections for the fermion mass
hierarchy, thereby keeping the pleasing picture of
technicolor \cite{TC} while avoiding the usual problems of
Extended Technicolor \cite{ETC}.
The phenomenological purpose of this excercise will be clear: the
discovery of e.g. strong $W_L W_L$ scattering
need not imply a usual 4-dimensional technicolor-like theory; it
could be the first signal of extra dimensions and strong gravitational
effects at the TeV scale.

As a simple toy example we begin with a theory with Weyl
``technifermions" $\Psi,\Psi^c$ of charge$+1,-1$ living on our three
brane, with a $U(1)_{TC}$
technicolor group living in the $(4+n)$ dimensional bulk. The point is
that $U(1)_{TC}$ has a dimensionful interaction strength, which in our
case is given by the only short distance scale available, the quantum
gravity
scale $M_s\sim$ TeV:
\begin{equation}
g^2_{TC} = \frac{h^2_{TC}}{M_s^n}
\end{equation}
where $h_{TC}$ is the dimensionless strength of $U(1)_{TC}$ at
$M_s$.
Like gravity itself, $U(1)_{TC}$ becomes strong near the UV cutoff
$\sim$ TeV, and the attractive
force it mediates between $\Psi,\Psi^c$ can trigger the condensate
$\langle \Psi \Psi^c \rangle$. The dynamics can be described by an
effective Nambu-Jona-Lasinio model, and the condensate will run to the
cutoff. This is just what is desired in our case,
since the cutoff is at the TeV scale.
To understand the physics better, consider the effective 4-fermion
operator induced between $\Psi,\Psi^c$ from tree-level exchange of the
$U(1)_{TC}$
gauge boson. Since $(4+n)$ dimensional momentum is not conseved, this
involves an integration:
\begin{equation}
{\cal O} \sim \int \frac{d^n \kappa}{(2\pi)^n} g^2_{TC} \bar{\Psi}
\bar{\sigma}^{\mu} \Psi \frac{1}{\kappa^2} \bar{\Psi^c}
\bar{\sigma}_{\mu} \Psi^c
\end{equation}
For $n>2$, the integral in the above is power divergent in the UV, and
must be cutoff at $M_s\sim 1$ TeV.
We then obtain a local, attractive four-fermion operator, whici upon
Fierzing becomes
\begin{equation}
{\cal O} \sim \frac{h_{TC}^2}{(2 \pi)^n M_s^2} \left(\Psi \Psi^c
\right)\left(\bar{\Psi} \bar{\Psi^c}\right)
\end{equation}
This is our effective NJL model, which triggers a $\langle
\Psi \Psi^c \rangle$
condensate for sufficiently large $h_{TC}$.

It is very easy to extend this analysis to a realistic model.
For aesthetic reasons,let the technifermions form one complete SM
generation;
it is of course possible
to choose a smaller technifermion sector. We will choose $U(1)_{TC}$ to
be $(B-L)$ for this technifamily.
This anomaly-free, and all the
$(B-L)$ invariant technifermion bilinears
\begin{equation}
(Q_{TC} U_{TC}^c), (Q_{TC} D_{TC}^c), (L_{TC} E_{TC}^c)
\end{equation}
have the correct Higgs quantum numbers under $SU(2)_L \times U(1)_Y$, so
there is no vacuum alignment problem.
Just as in our toy example, $U(1)_{TC}$ becomes strong in the
UV and forces a condensate to form. Of course all condensate channels
are
desirable in our case, however, $(LE^c)$ is the naive most attractive
channel and likely condenses first, triggering electroweak breaking.
Notice that since we do not need a non-abelian TC group, the
technifermion content can be at least twice as small as the usual
scenario with the minimal
$SU(2)_{TC}$ group, improving the situation with the $S$ parameter.
Furthermore, this theory is manifestly non-QCD like, so the usual
estimate of the
$S$ paramter in QCD-like theories does not apply.

Having used the bulk technicolor for dynamical symmetry breaking, we can
use all the ideas of the previous sections for generating small Yukawa
couplings
by simply replacing $H$ with $(\Psi \Psi^c)_{TC}$. We do not need
Extended Technicolor: the theory can still have a flavor symmetry
(under which the technifermions may or may not be charged), and the
theory above the TeV scale can naturally
generate TeV suppressed operators linking e.g. $LE^c (\Psi \Psi^c)_{TC}$
to the bulk messenger field $\chi$.
Of course, a detailed analysis of precision electroweak obsevables must
be made
to assess the viability of this particular model.
We do not perform this analysis here. We simply wish to point out that
there could be intrinsically higher-dimensional mechansims
foe electroweak symmetry breaking without a fundamental Higgs,
which in combination with our previous ideas for flavor,
can avoid the myriad of FCNC problems associated with Extended
Technicolor.

\section{Discussion and conclusions}
Solving the hierarchy problem by bringing the scale of quantum
gravity down to the TeV scale, in the presence of large new spatial
dimensions,
destroys the desert in short distance scales between the Weak and Planck
scales where many
mechanisms and phenomena, such as the origin of flavor, neutrino
masses and the longevity of the proton, enjoy their usual home.
In this paper we have shown that, instead of residing in the
ultra-short distances between $10^{-17}$cm and $10^{-33}$cm, these
phenomena can find a natural home in the large space of the extra
dimensions.

Our main focus has been to try to understand the origin of flavor
hierarchies at the TeV scale, while avoiding the usual
flavor-changing problems. We accomplish this by supposing that the
theory admits a flavor symmetry which prohibits the light
generation Yukawa couplings in the flavor symmetric limit. These
flavor symmetries are broken with $O(1)$ strength, but on distant
branes. This breaking acts as a source which ``shines" a bulk
scalar field $\chi$, charged under the flavor symmetry. The
smallness of the Yukawa couplings follows from the
small intensity of the $\chi$ vev shined on our brane, due to the
distance between the branes, giving an exponential suppression if $\chi$
is massive, and a power-law suppression if it is (nearly)
massless. The origin of the different hierarchies is then reduced
to the question of determining the inter-brane separations.
We did not address this dynamical problem in detail, but
remark that, for exponential or even sufficiently high power
suppressions, the branes need not be more separated by more than $O(10)$
times the fundamental length scale in order to span the observed
range of Yukawa couplings. The flavor symmetry must be sufficiently
powerful
to forbid dangerous FCNC operators even after it is broken.
We presented an explicit model based on the maximal $U(3)^5$
flavor symmetry where essentially the usual GIM mechanism explains
the absence of large FCNC effects.

These ideas for generating small couplings are completely generic and do
not
depend on a low quantum gravity scale. They can be used to explain
small neutrino masses and dynamical SUSY breaking in more
conventional theories with a high Planck scale.
We also re-iterated that proton longevity can result from gauged
baryon number in the bulk, broken on a far away brane. Another
interesting possibility for phenomenology from higher dimensions is
to do away with a fundamental Higgs on our brane, and trigger
electroweak breaking by technicolor dynamics in the bulk. Combined
with our other ideas for generating flavor, this can avoid the
usual problems of extended technicolor.

There are a number of dramatic experimental signals associated
with the various mechanisms suggested in this paper. For any value of
the fundamental scale
$M_s$, if the $\chi$ fields are very light with Compton wavelengths
between 1 micron
to a millimeter, their exchange gives rise to an attractive,
isotope dependent force $10^6$ times stronger than gravity at
sub-millimeter distances, and can not be missed by the upcoming
measurements of sub-millimeter gravitational strength forces.
Light gauge fields in the bulk give rise to similar strength {\it
repulsive} forces. If the $\chi's$ are heavier than about a GeV
but lighter than physical Higgs particles, they can be produced in
novel Higgs decays to light generation fermions + $\chi$, with a
width possibly comparable or exceeding the usual $b \bar{b}$ final
state. Finally, both bulk gauge fields and $\chi's$ can be
produced at colliders, leading to $\gamma + $ missing energy or
jet + missing energy signals similar to those from bulk graviton
production, with comparable or much larger rates.

\section*{Acknowledgements}
It is a pleasure to thank Alex Kagan, Jim Wells and especially Lawrence
Hall for
valuable discussions. Gia Dvali has informed us of some complementary
work
addressing flavor physics in the context of theories with low
quantum gravity scales \cite{Gia}. We thank him for discussions.
The work of NAH is supported by the Department of Energy under contract
DE-AC03-76SF00515. The work of SD is supported in part
by NSF grant PHY-9219345-004.

\def\simlt{\stackrel{<}{{}_\sim}}
\def\simgt{\stackrel{>}{{}_\sim}}
\newcommand{\cm}{Commun.\ Math.\ Phys.~}
\newcommand{\prl}{Phys.\ Rev.\ Lett.~}
\newcommand{\pr}{Phys.\ Rev.\ D~}
\newcommand{\pl}{Phys.\ Lett.\ B~}
\newcommand{\ibar}{\bar{\imath}}
\newcommand{\jbar}{\bar{\jmath}}
\newcommand{\np}{Nucl.\ Phys.\ B~}
\newcommand{\be}{\begin{equation}}
\newcommand{\en}{\end{equation}}
\newcommand{\ba}{\begin{eqnarray}}
\newcommand{\ea}{\end{eqnarray}}
\newcommand{\aG}{\alpha_G}

\def\pl#1#2#3{{\it Phys. Lett. }{\bf B#1~}(19#2)~#3}
\def\zp#1#2#3{{\it Z. Phys. }{\bf C#1~}(19#2)~#3}
\def\prl#1#2#3{{\it Phys. Rev. Lett. }{\bf #1~}(19#2)~#3}
\def\rmp#1#2#3{{\it Rev. Mod. Phys. }{\bf #1~}(19#2)~#3}
\def\prep#1#2#3{{\it Phys. Rep. }{\bf #1~}(19#2)~#3}
\def\pr#1#2#3{{\it Phys. Rev. }{\bf D#1~}(19#2)~#3}
\def\np#1#2#3{{\it Nucl. Phys. }{\bf B#1~}(19#2)~#3}
\def\mpl#1#2#3{{\it Mod. Phys. Lett. }{\bf #1~}(19#2)~#3}
\def\arnps#1#2#3{{\it Annu. Rev. Nucl. Part. Sci. }{\bf #1~}(19#2)~#3}
\def\sjnp#1#2#3{{\it Sov. J. Nucl. Phys. }{\bf #1~}(19#2)~#3}
\def\jetp#1#2#3{{\it JETP Lett. }{\bf #1~}(19#2)~#3}
\def\app#1#2#3{{\it Acta Phys. Polon. }{\bf #1~}(19#2)~#3}
\def\rnc#1#2#3{{\it Riv. Nuovo Cim. }{\bf #1~}(19#2)~#3}
\def\ap#1#2#3{{\it Ann. Phys. }{\bf #1~}(19#2)~#3}
\def\ptp#1#2#3{{\it Prog. Theor. Phys. }{\bf #1~}(19#2)~#3}

\end{document}